\documentclass[preprint]{sigplanconf}

\usepackage{amsmath}
\usepackage{listings}
\usepackage{booktabs}
\usepackage{courier}
\usepackage{xspace}
\usepackage{url}

\def\systemname#1{\textsf{#1}\xspace}
\newcommand{\HOLLight}{\systemname{HOL Light}}
\newcommand{\Prolog}{\systemname{Prolog}}
\newcommand{\leanCoP}{\systemname{leanCoP}}
\newcommand{\MESON}{\systemname{MESON}}
\newcommand{\leanTAP}{\systemname{leanTAP}}

\newcommand{\HOL}{\systemname{HOL}}
\newcommand{\ProverNine}{\systemname{Prover9}}

\newcommand{\Otter}{\systemname{Otter}}
\newcommand{\Isabelle}{\systemname{Isabelle}}
\newcommand{\Sledgehammer}{\systemname{Sledgehammer}}

\newcommand{\Mizar}{\systemname{Mizar}}

\newcommand{\Vampire}{\systemname{Vampire}}

\newcommand{\E}{\systemname{E}}

\newcommand{\SPASS}{\systemname{SPASS}}

\newcommand{\OCaml}{\systemname{OCaml}}

\newcommand{\Metis}{\systemname{Metis}}

\newcommand{\HH}{\systemname{HOL\hspace{-.5mm}\textsc{\raisebox{.5mm}{\scriptsize y}}Hammer}}

\newcommand{\Owner}[1]{} %

\newcounter{def1}

\newtheorem{definition}[def1]{Definition}

\newtheorem{theorem}{Theorem}[section]
\newtheorem{lemma}[theorem]{Lemma}

\begin{document}

\setlength{\pdfpageheight}{\paperheight}
\setlength{\pdfpagewidth}{\paperwidth}

\conferenceinfo{CONF 'yy}{Month d--d, 20yy, City, ST, Country}
\copyrightyear{20yy}
\copyrightdata{978-1-nnnn-nnnn-n/yy/mm}
\doi{nnnnnnn.nnnnnnn}

%
%

%
                                  %

%
                                  %
                                  %

%\titlebanner{Preprint submitted to CPP 2015}        %
%\preprintfooter{short description of paper}   %

\title{Certified Connection Tableaux Proofs for HOL Light and TPTP}

\authorinfo{Cezary Kaliszyk}
           {University of Innsbruck}
           {cezary.kaliszyk@uibk.ac.at}
\authorinfo{Josef Urban}
           {Radboud University Nijmegen}
           {josef.urban@gmail.com}
\authorinfo{Ji\v{r}\'i Vysko\v{c}il}
           {Czech Technical University in Prague}
           {jiri.vyskocil@gmail.com}

\maketitle

\begin{abstract}
In the recent years, the \Metis prover based on ordered paramodulation and model elimination
has replaced the earlier built-in methods for general-purpose proof automation in \HOL4 and \Isabelle/\HOL.
In the annual CASC competition, the \leanCoP system based on connection tableaux
has however performed better than \Metis.
In this paper we show how the \leanCoP's core algorithm can be
implemented inside \HOLLight. \leanCoP's flagship feature, namely its
minimalistic core, results in a very simple proof system. This plays
a crucial role in extending the \MESON proof reconstruction mechanism
to connection tableaux proofs, providing an
implementation of \leanCoP that certifies its proofs. We discuss the
differences between our direct implementation using an explicit \Prolog
stack, to the continuation passing implementation of \MESON present in
\HOLLight and compare their performance on all core \HOLLight goals.
The resulting prover can be also used as a general purpose TPTP prover.
We compare its performance against the resolution based \Metis
on TPTP and other interesting datasets.
\end{abstract}

\category{CR-number}{subcategory}{third-level}


\section{Introduction and Related Work}
\label{Introduction}
\Owner{Josef}

The \leanCoP \cite{OB03} automated theorem prover (ATP) has an
unusually good ratio of performance to its implementation size. %
While its core algorithm fits on some twenty lines of \Prolog,
starting with CASC-21~\cite{Sutcliffe08} it has regularly beaten
\Otter~\cite{MW97} and \Metis~\cite{hurd2003d} in the FOF division of
the CASC ATP competitions. In 2014, \leanCoP solved 158 FOF problems
in
CASC-J7,\footnote{\url{http://www.cs.miami.edu/~tptp/CASC/J7/WWWFiles/DivisionSummary1.html}}
while \ProverNine solved 95 problems. On the large-theory (chainy)
division of the MPTP Challenge
benchmark\footnote{\url{http://www.cs.miami.edu/~tptp/MPTPChallenge/}},
\leanCoP's goal-directed calculus beats also \SPASS 2.2~\cite{Weidenbach+99},
and its further AI-style strengthening by integrating into \leanCoP's simple core learning-based guidance trained on such larger ITP corpora 
is an interesting possibility~\cite{UrbanVS11}.

Compact ATP calculi such as \leanTAP~\cite{BeckertP95} and
\MESON~\cite{Loveland68} have been used for some time in
\Isabelle~\cite{Paulson99,Paulson97} and \HOL{}s~\cite{Har96} as general
first-order automation tactics for discharging goals that are already
simple enough.  With the arrival of large-theory ``hammer''
linkups~\cite{sledgehammer10,hhmcs,KaliszykU13b,KuhlweinBKU13} between ITPs,
state-of-the-art ATPs such as \Vampire~\cite{Vampire} and \E~\cite{Schulz13}, and premise selection methods~\cite{KuhlweinLTUH12}, such tactics
also became used as a relatively cheap method for reconstructing the
(minimized) proofs found by the stronger ATPs. In particular, Hurd's \Metis has
been adopted as the main proof reconstruction tool used by \Isabelle's
\Sledgehammer linkup~\cite{PaulsonS07,sledgehammer}, while Harrison's
version of \MESON could reconstruct in 1 second about 80\% of the
minimized proofs found by \E in the first experiments with the \HH
linkup~\cite{proch}.

Since \HOLLight already contains a lot of the necessary infrastructure
for \Prolog-style proof search and its reconstruction, integrating
\leanCoP into \HOLLight in a similar way as \MESON should not be too
costly, while it could lead to interesting strengthening of \HOLLight's
first-order automation and proof reconstruction methods.
In this paper we describe how
this was done, resulting in an \OCaml implementation of \leanCoP and a
general \leanCoP first-order tactic in \HOLLight. We compare their
performance with \MESON, \Metis and the \Prolog version of \leanCoP in
several scenarios, showing quite significant improvements over \MESON
and \Metis.

\section{leanCoP and Its Calculus}
\label{Leancop}
\Owner{Jirka}

\leanCoP is an automated theorem prover for classical first-order
logic based on a compact \Prolog implementation of the clausal connection
(tableaux) calculus \cite{OB03,LetzS01} with several
simple strategies that significantly reduce the search space on many
problems. In contrast to saturation-based calculi used in most of
the state-of-the-art ATPs (\E , \Vampire, etc.), connection calculi
implement goal-oriented proof search. Their main inference step connects
a literal on the current path to a new literal with the same predicate
symbol but different polarity. The formal definition (derived from Otten~\cite{DBLP:journals/aicom/Otten10})
of the particular \emph{connection
calculus} relevant in \leanCoP 
is as follows:

\begin{definition}{[}Connection calculus{]} 
\normalfont
The axiom and rules of
the \emph{connection calculus} are given in Figure \ref{fig:Calculus}.
The words of the calculus are tuples $"C,M,Path"$ where the clause
$C$ is the \emph{open subgoal}, $M$ is a set of clauses in disjunctive
normal form (DNF) transformed from \emph{axioms} $\land$ \emph{conjecture}
with added nullary predicate $\sharp$\footnote{We suppose that $\sharp$ is 
a new predicate that does not occur anywhere in \emph{axioms} and \emph{conjecture}} 
to all positive clauses\footnote{Thus by default all positive clauses are
used as possible start clauses.}, and the \emph{active path} $Path$
is a subset of a path through $M$ . In the rules of the calculus
$C,C'$ and $C''$ are clauses, $\sigma$ is a term substitution,
and ${L_{1},L_{2}}$ is a \emph{connection} with $\sigma(L_{1})=\sigma(\overline{L_{2}})$.
The rules of the calculus are applied in an analytic (i.e. bottom-up)
way. The term substitution $\sigma$ is applied to the whole derivation.
\end{definition} The connection calculus is correct and complete~\cite{LetzS01}
in the following sense: A first-order formula $M$ in clausal form
is valid iff there is a connection proof for $"\neg\sharp,M,\{\}"$,
i.e., a derivation for $"\neg\sharp,M,\{\}"$ in the connection calculus
so that all leaves are axioms.
The \Prolog predicate \texttt{prove/5}
implements the axiom, the
reduction and the extension rule of the basic connection calculus in \leanCoP:
\begin{lstlisting}[language=Prolog, numbers=left, keepspaces]
%
prove([Lit|Cla],Path,PathLim,Lem,Set) :-
	%
	(-NegLit=Lit;-Lit=NegLit) ->
	( %
	  %
	  member(NegL,Path), 
	  unify_with_occurs_check(NegL,NegLit)
	;
	  lit(NegLit,NegL,Cla1,Grnd1),
	  unify_with_occurs_check(NegL,NegLit),
	  %
	  %
	  prove(Cla1,[Lit|Path],PathLim,Lem,Set)
	),
	%
	prove(Cla,Path,PathLim,Lem,Set).
prove([],_,_,_,_).
\end{lstlisting}
The tuple ``$C,M,Path$''
in connection calculus is here represented as
follows:

\begin{itemize}
\item $C$ representing the open subgoal is a \Prolog list \texttt{Cla};
\item the active path \emph{Path} is a \Prolog list \texttt{Path};
\item \emph{M} is written into \Prolog's database before the actual proof search
starts in a such way that for every clause $C\in M$ and for every
literal $C\in M$ the fact \texttt{lit(Indexing\_L,L,C1,Grnd)} is
stored, where \texttt{C1}=\emph{C}\textbackslash{}\{\texttt{L}\} and
\texttt{Grnd} is \texttt{g} if \emph{C} is ground, otherwise \texttt{Grnd}
is \texttt{n}. \texttt{Indexing\_L} is same as \texttt{L} modulo all
its variables which are fresh (there is no twice or more occurrences
in \texttt{Indexing\_L}) everywhere in \texttt{Indexing\_L} and it
is used for fast finding the right fact in database without affecting
the logically correct \texttt{L} by standard \Prolog unification
without occurs check.
\item Atoms are represented by \Prolog atoms, negation
by ``$-$''.
\item The substitution $\sigma$ is stored implicitly by
\Prolog.
\end{itemize}
\texttt{PathLim} is the current limit
used for iterative deepening, \texttt{Lem} is the list of usable (previously
derived) lemmas, \texttt{Set} a list of options, and \texttt{Proof}
is the resulting proof. 
This predicate succeeds (using iterative deepening) iff
there is a connection proof for the tuple represented by \texttt{Cla},
the DNF representation of the problem stored in \Prolog's database using the
\texttt{lit} predicate, and a \texttt{Path} with $|$\texttt{Path}$|<$\texttt{PathLim}
where \texttt{PathLim} is the maximum size of the active \texttt{Path}.
The predicate works as follows: %

Line 18 implements the axiom, line 4 calculates the complement of the
first literal \texttt{Lit} in \texttt{Cla}, which is used as the principal
literal for the next reduction or extension step. The reduction rule
is implemented in lines 7, 8 and 17.
At line 7 and 8 it is checked whether the active path \texttt{Path} contains
a literal \texttt{NegL} that unifies with the complement \texttt{NegLit}
of the principal literal \texttt{Lit}. In this case the alternative
lines after the semicolon are skipped and the proof search for the
premise of the reduction rule is invoked in line 17.
The extension rule is implemented in lines 10, 11, 14 and 17. 
Lines 10 and 11 are used to find a clause
that contains the complement \texttt{NegLit} of the principal literal
\texttt{Lit}. \texttt{Cla1} is the remaining set of literals of the
selected clause and the new open subgoal of the left premise. The proof
search for the left premise of the extension rule, in which the active
path \texttt{Path} is extended by the principal literal \texttt{Lit},
is invoked in line 14,
and if successful, we again continue on line 17.

\begin{figure}
\begin{description}
\item [{axiom:}] $\dfrac{}{\{\},M,Path}$ 
\item [{reduction~rule:}] $\dfrac{C,M,Path\cup\{L_{2}\}}{C\cup\{L_{1}\},M,Path\cup\{L_{2}\}}$
\\
 where there exists a unification substitution $\sigma$ such that
$\sigma(L_{1})=\sigma(\overline{L_{2}})$ 
\item [{extension~rule:}] $\dfrac{C'\setminus\{L_{2}\},M,Path\cup\{L_{1}\}\qquad C,M,Path}{C\cup\{L_{1}\},M,Path}$
\\
 where $C'$is a fresh copy of some $C''\in M$ such that $L_{2}\in C'$and
$\sigma(L_{1})=\sigma(\overline{L_{2}})$ where $\sigma$ is unification
substitution. 
\end{description}
Note that the $\sigma$ used in the \emph{reduction} and \emph{extension}
rules must be applied on all literals in all derivations except the
literals in the set $M$ because these literals are not affected by
any substitution $\sigma$.

\caption{\label{fig:Calculus}The basic clause connection calculus used in
\leanCoP. }
\end{figure}

Compared to standard tableaux or sequent calculi, connection calculi
are not confluent%
\footnote{Bad choice of \emph{connection} might end up in dead end%
}. To achieve completeness, an extensive use of backtracking is required.
\leanCoP uses two simple incomplete strategies (namely options \texttt{scut}
and \texttt{cut}) for restricting backtracking %
that significantly reduces the search space \cite{DBLP:journals/aicom/Otten10}
without affecting the ability to find proofs in most tested cases
(see Section \ref{Experiments}).

Another major problem in connection calculi is the integration of
equality. The paramodulation method that is widely used in saturation-based
ATPs is not complete for goal-oriented approach of connection calculi.
Therefore equality in \leanCoP and similar ATPs is usually managed by
adding the axioms of equality (reflexivity, symmetry, transitivity
and substitutivity).

To obtain the
clausal form, \leanCoP uses its own implementation of %
clausifier introducing definitions (the \texttt{def} option), which seems to perform better with the
\leanCoP's core prover than other standard clausifiers (TPTP2X
using the option \texttt{-t clausify:tptp}, FLOTTER and \E) or direct
transformation into clausal form (\texttt{nodef} option in \leanCoP)
\cite{DBLP:journals/aicom/Otten10}. In the following subsections, we summarize several further methods used by \leanCoP that improve its performance.

\subsection{Iterative deepening}

\Prolog uses a simple incomplete depth-first
search strategy to explore the search space. This kind of incompleteness would result
in a calculus that hardly proves any formula. In order
to obtain a complete proof search in the connection
calculus, iterative deepening on the proof depth, i.e. the
size of the active path, is performed. It is achieved by
inserting the following lines into the code:\\
\begin{minipage}[c]{0.57\textwidth}%
\begin{lstlisting}[language=Prolog]
(12)( Grnd1=g -> true ; length(Path,K),
			 K<PathLim -> true ;
(13) \+ pathlim -> assert(pathlim), fail ),
\end{lstlisting}
\end{minipage}
and the whole prover runs in the following
iterative sense starting from \texttt{PathLimit}$=1$:\\
\begin{minipage}[c]{0.57\textwidth}%
\begin{lstlisting}[language=Prolog]
prove(PathLim,Set) :-
	retract(pathlim) ->
	PathLim1 is PathLim+1,
	prove(PathLim1,Set).
\end{lstlisting}
\end{minipage}
When the extension rule is applied and the new
clause is not ground, i.e. it does not contain any variable,
it is checked whether the size \texttt{K} of the active path
exceeds the current path limit \texttt{PathLim} (line 12).
In this case the dynamic predicate \texttt{pathlim/0} is written into
the \Prolog's database (line 13) indicating the need to increase
the path limit if the proof search with the current
path limit fails. If the proof search fails and the
predicate \texttt{pathlim} can be found in the database, 
then \texttt{PathLim} is increased by one and the proof search
starts again.

\subsection{Regularity Condition Optimization}
\begin{definition}
\normalfont
A connection proof is \emph{regular}
iff no literal occurs more than once in the active path.
\end{definition}
Since the active path corresponds to the set of literals
in a branch in the connection tableau representation,
a connection tableau proof is regular if in the current
branch no literal occurs more than
once. The regularity condition is integrated
into the connection calculus in Figure \ref{fig:Calculus} by imposing
the following restriction on the reduction and
extension rule: $\forall L'\in C\cup\{L_{1}\}:\sigma(L')\notin\sigma(Path)$\\
\begin{lemma}
A formula $M$ in clausal form described above
is valid iff there is a regular connection proof for
$"\neg\sharp,M,\{\}"$
\end{lemma}
Regularity is correct, since it only imposes a restriction
on the applicability of the reduction and extension
rules. The completeness proof can be found in \cite{OB03,LetzS01}.\\
The regularity condition must be checked
whenever the reduction, extension or lemma rule is applied.
The substitution $\sigma$ is not modified, i.e. the regularity
condition is satisfied if the open subgoal does not
contain a literal that is syntactically identical with a literal
in the active path. This is implemented by inserting
the following line into the code:\\
\begin{minipage}[c]{0.57\textwidth}%
\begin{lstlisting}[language=Prolog]
(3) \+ (member(LitC,[Lit|Cla]), 
	member(LitP,Path), 
	LitC==LitP),
\end{lstlisting}
\end{minipage}
The \Prolog predicate \texttt{\textbackslash +} \emph{Goal} succeeds only if \emph{Goal}
cannot be proven. In line 3 the corresponding \emph{Goal} succeeds
if the open subgoal \texttt{[Lit|Cla]} contains a literal
\texttt{LitC} that is syntactically identical (built-in predicate \texttt{==/2} in \Prolog)
with a literal \texttt{LitP} in the active path \texttt{Path}. The built-in
predicate \texttt{member/2} is used to enumerate all elements
of a list. 

\subsection{Lemmata optimization}
The set of lemmata is represented by the list \texttt{Lem}.
The lemma rule is implemented
by inserting the following lines:\\
\begin{minipage}[c]{0.57\textwidth}%
\begin{lstlisting}[language=Prolog]
(5) ( member(LitL,Lem), Lit==LitL
(6) ;
\end{lstlisting}
\end{minipage}
In order to apply the lemma rule, the substitution $\sigma$ is
not modified, i.e. the lemma rule is only applied if the
list of lemmata \texttt{Lem} contains a literal \texttt{LitL} that is syntactically
identical with the literal \texttt{Lit}. Furthermore,
the literal \texttt{Lit} is added to the list \texttt{Lem} of lemmata in
the (left) premise of the reduction and extension rule
by adapting the following line:\\
\begin{minipage}[c]{0.57\textwidth}%
\begin{lstlisting}[language=Prolog]
(15) prove(Cla,Path,PathLim,[Lit|Lem],Set).
\end{lstlisting}
\end{minipage}
In the resulting implementation, the lemma rule is
applied before the reduction and extension rules.

\subsection{Restricted backtracking}
In \Prolog the
cut (!) is used to cut off alternative solutions when
\Prolog tries to prove a goal. The \Prolog cut is a built-in
predicate, which succeeds immediately when first encountered
as a goal. Any attempt to re-satisfy the cut
fails for the parent goal, i.e. other alternative choices
are discarded that have been made from the point when
the parent goal was invoked. Consequently, \emph{restricted
backtracking} is achieved by inserting a Prolog cut after
the lemma, reduction, or extension rule is applied. It
is implemented by inserting the following line into the
code:\\
\begin{minipage}[c]{0.57\textwidth}%
\begin{lstlisting}[language=Prolog]
(16) ( member(cut,Set) -> ! ; true ),
\end{lstlisting}
\end{minipage}
Restricted backtracking is switched on if the list
\texttt{Set} contains the option \texttt{cut}.\\
The \emph{restricted start step} is used if the list \texttt{Set} includes
the option \texttt{scut}. In this case only the first matching clause
to starting $\neg\sharp$ literal is used.\\
Restricted backtracking and restricted start step lead to an incomplete
proof search. In order to regain completeness, these strategies can be
switched off when the search reaches a certain path limit.
If the list \texttt{Set} contains the option \texttt{comp(}\emph{Limit}\texttt{)}, 
where \emph{Limit} is a natural number,
the proof search is stopped and started again without using incomplete search
strategies.

\section{\OCaml Implementation}
\label{Implementation}
\Owner{Cezary}

In this section, we first discuss our implementation\footnote{Available online at \url{http://cl-informatik.uibk.ac.at/users/cek/cpp15}}  of \leanCoP in \OCaml
and its integration in \HOLLight: the transformation of the higher-order
goal to first order and the proof reconstruction. After that we compare our
implementation to Harrison's implementation of \MESON.

\subsection{\leanCoP in \OCaml}

Otten's implementation of \leanCoP uses the \Prolog search, backtracking, and indexing
mechanisms to implement the connection tableaux proof search. This is a variation of the general idea of 
using the ``Prolog technology theorem prover'' (PTTP) proposed by Stickel~\cite{Stickel88}, in which connection tableaux takes a number of advantages from its similarity to \Prolog,

In order to implement an equivalent program
in a functional programming language, one needs to use either an explicit stack for keeping track of the current proof state (including the trail of variable bindings),
or the continuation passing style. We choose to do the former, namely we add an explicit
\texttt{todo} (stack), \texttt{subst} (trail) and \texttt{off} (offset in the trail) arguments to the main \texttt{prove} function.
The stack keeps a list of tuples
that are given as arguments to the recursive invocations of \texttt{prove}
, whose full \OCaml declaration (taking the open subgoal as its last argument) then looks as follows:\\
\begin{minipage}{.50\textwidth}
\begin{lstlisting}{Name}
let rec lprove off subst path limit lemmas todo = function
  [] -> begin ... end 
  | ((lit1 :: rest_clause) as clause) -> ... ;;
\end{lstlisting}
\end{minipage}

The function %
performs
the proof search to the given depth, and if a proof has not been found, it returns
the unit. It takes special attention to traverse the tree in the same
order as the \Prolog version. In particular, when the global option "cut" (restricting backtracking) is off, it performs
all the backtrackings
explicitly, while if "cut" is on, the parts of backtracking avoided in
\Prolog are also omitted.
When a proof is found, the exception 'Solved' is raised: no further
search is performed
and the function exits with this exception.

The \OCaml version and the \Prolog version (simplified and with symbols renamed for clarity of comparison) are displayed together in
Fig.~\ref{listing}. The algorithm proceeds as follows:

\begin{figure*}[htb!]
\begin{minipage}{.01\textwidth}
\begin{lstlisting}{Name}
1

2


3

4


5



6










7
\end{lstlisting}
\end{minipage}
\begin{minipage}{.57\textwidth}
\begin{lstlisting}[language=ml]{Name}
let rec prove path lim lem stack = function (lit :: cla) ->

  if not ((exists (fun litp -> exists (substeq litp) path))
    (lit :: cla)) then (

  let neglit = negate lit in

  if not (exists (substeq lit) lem &&
     (prove path lim lem stack cla; cut)) then (

  if not (fold_left (fun sf plit -> if sf then true else
    try (unify_lit neglit plit; prove path lim (lit :: lem)
    stack cla; cut) with Unify -> sf) false path) then (

  let iter_fun (lit2, cla2, ground) = if lim > 0 || ground then
    try let cla1 = unify_rename (snd lit) (lit2, cla2) in
    prove (lit :: path) (lim - 1) lem ((if cut then lim else -1),
    path, lim, lit :: lem, cla) :: stack) cla1 with Unify -> () in
  try iter iter_fun (try assoc neglit lits with Not_found -> [])
  with Cut n -> if n = lim then () else raise Cut n)))

| [] -> match stack with
      (ct, path, lim, lem, cla) :: t ->
            prove path lim lem t cla; if ct > 0 raise (Cut ct)

    | [] -> raise Solved;;
\end{lstlisting}
\end{minipage}
\begin{minipage}{.41\textwidth}
\begin{lstlisting}[language=Prolog]{Name}
prove([Lit|Cla],Path,PathLim,Lem,Set) :-

  \+ (member(LitC,[Lit|Cla]), member(LitP,Path),
     LitC==LitP),

  (-NegLit=Lit;-Lit=NegLit) -> (

      member(LitL,Lem), Lit==LitL
      ;

      member(NegL,Path),
      unify_with_occurs_check(NegL,NegLit)
      ;

      lit(NegLit,NegL,Cla1,Grnd1),
      unify_with_occurs_check(NegL,NegLit),
        ( Grnd1=g -> true ;
          length(Path,K), K<PathLim -> true ;
          \+ pathlim -> assert(pathlim), fail ),
      prove(Cla1,[Lit|Path],PathLim,Lem,Set)
    ), ( member(cut,Set) -> ! ; true ),


    prove(Cla,Path,PathLim,[Lit|Lem],Set).

prove([],_,_,_,_,[]).
\end{lstlisting}
\end{minipage}
\caption{\label{listing}The simplified \OCaml and \Prolog code side by side. The explicit trail argument and
  the computation of the resulting proof have been omitted for clarity, and some symbols were renamed to better correspond to each other.
White-space
  and order of clauses has been modified to exemplify corresponding parts of the two implementations.
Function \texttt{substeq} checks equality under the current context of variable bindings. Note that the last-but-one line in the \Prolog code was merged into each of the three cases in the \OCaml code.
See the function \texttt{lprove} in file \texttt{leancop.ml} on our web page for the actual (non-simplified) \OCaml code.
}

\end{figure*}

\begin{enumerate}
\item If nonempty, decompose the current open subgoal into the first literal \texttt{lit} and the rest \texttt{cla}.
\item Check for intersection between the current open subgoal and \texttt{path}.
\item Compute the negation of \texttt{lit}.
\item Check if \texttt{lit} is among the lemmas \texttt{lem}, if so try to prove
  \texttt{cla}. If \texttt{cut} is set, no other options are tried.
\item For each literal on the \texttt{path}, if \texttt{neglit} unifies with it, try to prove
  \texttt{cla}. If the unification succeeded and \texttt{cut} is set, no other
  options are tried.
\item For each clause in the matrix, try to find a literal that unifies
  with \texttt{neglit}, and then try to prove the rest of the newly created subgoal and the rest of the
  current open subgoal. If the unification and the first proof succeeded and \texttt{cut}
  is set, no other options are tried.
\item When the current open subgoal is empty, the subproof is finished (the axiom rule). 
\end{enumerate}

In Otten's implementation, the behaviour of the program with \texttt{cut} set is
enabled by the use of the \Prolog cut (\texttt{!}). Implementing it in \OCaml
amounts to a different mechanism in each of the three cases. In point 4 in the
enumeration above,
given that a single lemma has been found, there is no need to check for other
lemmas. Therefore a simple \texttt{List.exists} call is sufficient to emulate this
behaviour in \OCaml. No backtracking over other possible occurrences of 
the lemma is needed here, and it is not necessary to add in this case the literal again into the list of lemmas 
as is done in the \Prolog code (last-but-one line).

In point 5, multiple literals on the path may unify with
different substitutions. We therefore use a list fold mechanism which changes
the value whenever the unification is successful and \texttt{cut} is set. In
point 6, we need to change our behaviour in between two successive calls to
\texttt{prove}. As the first call takes the arguments to the second call on the
stack, we additionally add a cut marker on the stack and handle an exception
that can be raised by the call on the stack.

Whenever the clause becomes empty, all the tuples in the stack list need
to be processed. For each tuple, the first component is the cut marker:
if it is set, the \texttt{Cut} exception is raised with a depth level. The
exception is handled only at the appropriate level. This directly corresponds
to the semantics of the cut operator in \Prolog~\cite{DBLP:conf/lopstr/StroderESGF11}.

What remains to be implemented is efficient equality checking and unification.
Since we want to integrate our mechanism in \HOLLight, we reuse the
first order logic representation used in the implementation of \HOLLight's \MESON
procedure: the substitutions are implemented as association lists, and applications
of substitutions are delayed until an actual equality check or a unification step.

\subsection{\leanCoP in \HOLLight}
In order to enable the \OCaml version of \leanCoP as a proof tactic and procedure in \HOLLight,
we first need to transform a HOL goal to a \leanCoP problem and when a proof has
been found we replay the proof in higher-order logic.
In order to transform a problem in higher-order logic to first-order logic
without equality, we mostly reuse the steps of the transformation already used
by \MESON, additionally ensuring that the conjecture is separated from the axioms
to preserve \leanCoP's goal-directed approach. The transformation starts by assuming
the duplicated copies of polymorphic theorems to match the existing assumptions. Next
the goal $axioms \to conjecture$ is transformed to $(axioms \land \sharp) \to (conjecture \land \sharp)$
with the help of a special symbol, which we define in HOL as: $\sharp = \top$.
Since the conjecture is refuted and the problem is converted to CNF, the only
positive occurrence of $\sharp$ is present in a singleton clause, and the negative
occurrences of $\sharp$ are present in every clause originating from the conjecture.
The CNF clauses directly correspond to the DNF used by the \Prolog version of \leanCoP. Since no steps
distinguish between the positivity of literals, the two can be used interchangeably
in the proof procedure. We start the FOL algorithm by trying to prove $\lnot \sharp$.

Since the final \leanCoP proof may include references to lemmas, the reconstruction cannot be performed
the same way as it is done in \MESON. There, a tree structure is used for finished
proofs. Each subgoal either closes the branch (the literal is a negation of a literal
already present on the path) or is a branch extension with a  (possibly empty) list
of subgoals. In \leanCoP, each subgoal can refer to previous subgoals, so the order
of the subgoals becomes important. We therefore flatten the tree to a list,
which needs to be traversed in a linear order to reconstruct the proof.

We define a type of proof steps, one for each proof step in the calculus. Each
application of a lemma step or path step constructs a proof step with
an appropriate first-order term. For an application of a tableaux
extension step we use Harrison's contrapositive mechanism: we store
the reference to the actual transformed HOL theorem whose conclusion
is a disjunction together with the number of the disjunct that got
resolved\footnote{Contrary to the name, the \HOLLight type \texttt{fol\_atom}
implements a literal: it is either positive or negative.}.

\begin{lstlisting}[language=ml]{OCaml}
type proof = Lem of fol_atom
           | Pat of fol_atom
           | Res of fol_atom * (int * thm);;
\end{lstlisting}

A list of such proof steps together with a final substitution and an initially
empty list of already proved lemmas are the complete input to the proof
reconstruction procedure. The reconstruction procedure always looks at the
first step on the list. First, a HOL term is constructed from the FOL term with
the final substitution applied. This step is straightforward, as it amounts to
reversing the mapping of variables and constants applied to transform the HOL CNF
to FOL CNF, with new names invented for new variables. Next, we analyze the kind of the step.
If the step is a path step, the theorem $tm \vdash tm$ is
returned, using the HOL \texttt{ASSUME} proof rule. If the step
is a lemma step, the theorem whose conclusion is equal to $tm$ is found on the list
of lemmas and returned. Finally, if the proof step is an extension step,
we first find the disjuncts of the HOL theorem in the proof step apart from the
one that got matched. We then fold over this list, at every step calling the
reconstruction function recursively with the remaining proof steps and the list of
lemmas extended by each of the calls. The result of the fold is the list of theorems
$[\vdash tm_1, \vdash tm_2, ..., \vdash tm_n]$ which gets matched with the contrapositive theorem
$\vdash tm_1 \land \ldots \land tm_n \to tm_g$ using the HOL proof rule \texttt{MATCH\_MP}
to obtain the theorem $\vdash tm_g$. Finally, by matching this theorem to the term $tm$
the theorem $\vdash tm$ is obtained.

As the reconstruction procedure traverses the list, it produces the theorem that
corresponds to the first goal, namely $... \vdash \lnot \sharp$. By unfolding
the definition of $\sharp$, we obtain $... \vdash \bot$ which concludes the
refutation proof.

\subsection{Comparison to \MESON}

The simplified \OCaml code of the core \HOLLight's \MESON algorithm as described in~\cite{HarrisonHandbook}
is as follows: %

\begin{lstlisting}[language=ml]{Name}
let rec mexpand rules ancestors g cont (env,n,k) =
  if n < 0 then failwith "Too deep" else 
    try tryfind (fun a -> cont (unify_literals 
				  env (g,negate a),n,k))
	  ancestors
    with Failure _ -> tryfind (fun rule -> 
      let (asm,c),k’ = renamerule k rule in
      itlist (mexpand rules (g::ancestors)) asm cont
	(unify_literals env (g,c),n - length asm,k’))
      rules;;

let puremeson fm =
  let cls = simpcnf(specialize(pnf fm)) in
  let rules = itlist ((@) ** contrapositives) cls [] in
  deepen (fun n -> mexpand rules [] False (fun x -> x) 
                           (undefined,n,0); n) 
    0;;
\end{lstlisting}

The toplevel \texttt{puremeson} function proceeds by turning the input
formula into a clausal form, making contrapositives (\texttt{rules}) from the
clauses, and then repeatedly calling the \texttt{mexpand} function
with these rules using iterative deepening over the number of nodes
permitted in the proof tree.

The \texttt{mexpand} function takes as its arguments the
\texttt{rules}, an (initially empty) list of goal \texttt{ancestors},
the goal \texttt{g} to prove (initially \texttt{False}, which was also
added to all-negative clauses when creating contrapositives), a
continuation function \texttt{cont} for solving the rest of the
subgoals (initially the identity), and a tuple consisting of the
current trail \texttt{env}, the number \texttt{n} of additional nodes
in the proof tree permitted, and a counter \texttt{k} for variable
renaming.

If the allowed node count is not negative, \texttt{mexpand} first
tries to unify the current goal with a negated ancestor, followed by
calling the current continuation (trying to solve the remaining goals)
with the extended trail. If all such unification/continuation attempts
fail (i.e., they throw Failure), an extension step is tried with all
rules.  This means that the head of a (renamed) rule is unified with
the goal \texttt{g}, the goal is appended to the ancestors and the
\texttt{mexpand} is called (again using list folding with the subsequently modified trail and continuation) for
all the assumptions of the rule, decreasing the allowed node count for
the recursive calls.

The full \HOLLight version of \MESON additionally uses a smarter
(divide-and-conquer) policy for the size limit, checks the goal for
being already among the ancestors, caches continuations, and uses
simple indexing. Below we enumerate some of the most important
differences between the \leanCoP algorithm and \MESON and their
implementations in \HOLLight. Their practical effect is measured in Section~\ref{Experiments}. 
\begin{itemize}
\item \leanCoP computes and uses lemmas. The literals that correspond to closed
  branches are stored in a list. Each call to the main \texttt{prove} function
  additionally looks for the first literal in the list of lemmas. This can
  cost a linear number of equality checks if no parts of the proof are reused,
  but it saves computations if there are repetitions.

\item Both algorithms use iterative deepening; however the depth and termination
  conditions are computed differently. 

\item \MESON is implemented in the continuation-passing-style, so it can use as an additional optimization caching of
  the continuations. %
If any continuations are
  repeated (at the same depth level), the subproof is not retried. Otten's
  \leanCoP uses a direct \Prolog implementation which cannot (without further tricks) do
  such repetition elimination. The implementation of \leanCoP in \OCaml behaves the same.

\item \leanCoP may use the cut after the lemma step, path step or successful branch
  closing in the extension step. Implementing this behaviour in \OCaml exactly
  requires multiple \texttt{Cut} exceptions -- one for each depth of the proof.

\item The checking for repetitions is done in a coarser way in \MESON
  than in \leanCoP, allowing \leanCoP to skip some work done by
  \MESON.

\item The search is started differently in \leanCoP and in
  \MESON. \leanCoP starts with a conjecture clause, which likely contributes to its relatively good performance on larger problems.

\end{itemize}

\section{\label{Experiments}Experimental Setup and Results}

For the experiments we use \HOLLight SVN version 199 (September 2014),
\Metis 2.3, and \leanCoP 2.1. Unless noted otherwise, the systems are run on a 48-core server with AMD
Opteron 6174 2.2 GHz CPUs, 320 GB RAM, and 0.5 MB L2 cache per CPU.
Each problem is always assigned one CPU.

The systems are compared on several benchmarks, corresponding to
different modes of use: goals coming from \HOLLight itself, the
general set of problems from the TPTP library, and the large-theory
problems extracted from \Mizar~\cite{Urban06}. The first set of
problems is important, however since these problems come from
\HOLLight itself, they are likely naturally biased towards \MESON. The
\Mizar problems come from the two MPTP-based benchmarks: the MPTP
Challenge and MPTP2078~\cite{abs-1108-3446}. These are large-theory
problems coming from a different ITP, hence they do not introduce the
implicit bias as the \HOLLight problems, while coming from a more
relevant application domain than the general TPTP problems.

For \HOLLight, we evaluate (with 5 second time limit) on two sets of problems. First, we look at
872 \MESON-solved \HOLLight goals that were made harder by removing
splitting. In this scenario the tactic is applied to a subgoal of a
proof, which is a bit similar to the Judgement-Day~\cite{sledgehammer}
evaluation used for \Isabelle/\Sledgehammer, where the goals are
however restricted to the solvable ones. Table~\ref{tab:subgoal} shows
the results.  Second, we evaluate on the top-level goals (with their
dependencies already minimized) that have been solved with the \HH
system~\cite{holyhammer}, i.e., by using the strongest available ATPs.
This set is important because tactics such as \MESON, \Metis and now
also \leanCoP can be tried as a first cheap method for reconstructing
the proofs found by the stronger ATPs. The results are shown in
Table~\ref{tab:toplevel}.  In both cases, the \OCaml implementation of
\leanCoP performs best, improving the \MESON's performance in the
first case by about 11\%, and improving on \Metis on the second set of
problems by about 45\%.

Table~\ref{tab:tptp} shows the results of the evaluation on all 7036
FOF problems coming from TPTP 6.0.0, using 10 second time limit. Here
the difference to \Metis is not so significant, probably because
\Metis implements ordered paramodulation, which is useful for many
TPTP problems containing equality. The improvement over \MESON is
about 17\%.  Table~\ref{tab:oldbushy} and Table~\ref{tab:oldchainy}
show the results on the small (heuristically minimized) and large
MPTP Challenge problems. The best version of the \OCaml implementation
of \leanCoP improves by 54\% on \Metis and by 90\% on \MESON on the
small problems, and by 88\% on \Metis and 100\% on \MESON on the large
problems. Here the goal directedness of \leanCoP is probably the main factor.

Finally, to get a comparison also with the best ATPs on a larger
ITP-oriented benchmark (using different hardware), we have done a 10s
evaluation of several systems on the newer MPTP2078 benchmark (used in
the 2012 CASC@Turing competition), see Table~\ref{tab:newbushy} and
Table~\ref{tab:newchainy}. The difference to \Metis and \MESON on the
small problems is still quite significant (40\% improvement over
\MESON), while on the large problems the goal-directedness again shows
even more (about 90\% improvement). While \Vampire's (version 2.6) SInE heuristic~\cite{HoderV11}
helps a lot on the larger problems~\cite{UrbanHV10}, the difference there between \E (1.8)
and our version of \leanCoP is not so great as one could imagine given
the several orders of magnitude difference in the size of their
implementations.

\begin{table}
  \centering
\begin{tabular}{lcc}\toprule
Prover & Theorem (\%) & Unique \\\midrule %
mlleancop-cut-comp & 759 (87.04) & 2 \\%& 0.186 & 141.20 & 0 (0.00) \\
mlleancop-nocut & 759 (87.04) & 2 \\%& 0.185 & 140.12 & 0 (0.00) \\
plleancop-cut & 752 (86.23) & 0 \\%& 0.180 & 135.12 & 0 (0.00) \\
plleancop-nc & 751 (86.12) & 0 \\%& 0.180 & 134.92 & 0 (0.00) \\
metis-23 & 708 (81.19) & 26 \\%& 0.217 & 153.82 & 0 (0.00) \\
meson & 683 (78.32) & 4 \\\midrule %
any & 832 (95.41) \\ \bottomrule %
\end{tabular}
  \caption{\label{tab:subgoal}Core HOL Light MESON calls without splitting (872 goals), 5sec per goal}
\end{table}

\begin{table}
  \centering
\begin{tabular}{lcc}\toprule
Prover & Theorem (\%) & Unique \\\midrule %
mlleancop-cut-comp & 1178 (75.70) & 12 \\%& 0.205 & 241.55 & 0 (0.00) \\
mlleancop-nocut & 1162 (74.67) & 0 \\ %
meson & 1110 (71.33) & 39 \\ %
plleancop-nc & 1085 (69.73) & 0 \\ %
plleancop-cut & 1084 (69.66) & 0 \\ %
metis-23 & 814 (52.31) & 16 \\ \midrule %
any & 1260 (80.97) \\ \bottomrule %
\end{tabular}
\caption{\label{tab:toplevel}HOL Light dependencies (1556 goals, 5sec)}
\end{table}

\begin{table}
  \centering
\begin{tabular}{lccccc}\toprule
Prover & Theorem (\%) & Unique \\ \midrule %
mlleancop-cut-conj & 1669 (23.72) & 73 \\ %
plleancop-cut-conj & 1648 (23.42) & 21 \\ %
plleancop-cut & 1622 (23.05) & 34 \\ %
mlleancop-cut & 1571 (22.32) & 9 \\ %
metis-23 & 1562 (22.20) & 261 \\ %
meson & 1430 (20.32) & 28 \\ %
plleancop-nocut & 1358 (19.30) & 25 \\ %
mlleancop-nocut & 1158 (16.45) & 3 \\ \midrule %
any & 2433 (34.57)\\ \bottomrule %
\end{tabular}
\caption{\label{tab:tptp}TPTP (7036 goals with at least one conjecture, 10sec)}
\end{table}

\begin{table}
  \centering
\begin{tabular}{lcccccc}\toprule
Prover & Theorem (\%) & Unique \\ %
pllean-cut-conj & 103 (40.87302) & 2 \\ %
pllean-cut & 99 (39.28571) & 8 \\ %
mlleancop-cut-conj & 91 (36.11111) & 2 \\ %
mlleancop-cut & 79 (31.34921) & 0 \\ %
mlleancop-nocut & 76 (30.15873) & 0 \\ %
pllean-nc & 62 (24.60317) & 1 \\ %
metis-23 & 59 (23.41270) & 3 \\ %
meson-infer & 48 (19.04762) & 0 \\ \midrule %
any & 124 (49.20635) \\ \bottomrule %
\end{tabular}
\caption{\label{tab:oldbushy} Bushy (small) MPTP Challenge problems (252 in total), 10sec)}
\end{table}

\begin{table}
  \centering
\begin{tabular}{lcccccc}\toprule
Prover & Theorem (\%) & Unique \\ %
pllean-cut-conj & 61 (24.20635) & 5 \\ %
mlleancop-cut-conj & 60 (23.80952) & 9 \\ %
pllean-cut & 57 (22.61905) & 4 \\ %
mlleancop-nocut & 47 (18.65079) & 0 \\ %
mlleancop-cut & 47 (18.65079) & 0 \\ %
metis-23 & 32 (12.69841) & 3 \\ %
meson-infer & 30 (11.90476) & 0 \\ %
pllean-nc & 26 (10.31746) & 0 \\ \midrule %
any & 83 (32.93651) \\ \bottomrule %
\end{tabular}
\caption{\label{tab:oldchainy} Chainy (large) MPTP Challenge problems (252 in total), 10sec)}
\end{table}

\begin{table}
\centering
\begin{tabular}{lc}\toprule
Prover & Theorem (\%) \\\midrule
\Vampire & 1198 (57.65) \\
e18 & 1022 (49.18) \\
mlleancop-cut-conj & 613 (29.49) \\
pllean-cut-conj & 597 (28.72) \\
metis-23 & 564 (27.14) \\
mlleancop-cut & 559 (26.90) \\
pllean-cut & 544 (26.17) \\
pllean-comp7 & 539 (25.93) \\
mlleancop-nocut & 521 (25.07) \\
pllean-nc & 454 (21.84) \\
meson-infer & 438 (21.07) \\
any & 1277 (61.45) \\\bottomrule
\end{tabular}
\caption{\label{tab:newbushy} Bushy (small) MPTP2078  problems (2078 in total), 10sec)}
\end{table}

\begin{table}
  \centering
\begin{tabular}{lc}\toprule
Prover & Theorem (\%) \\\midrule
\Vampire & 634 (30.51) \\
e18 & 317 (15.25) \\
mlleancop-cut-conj & 243 (11.69) \\
pllean-cut-conj & 196 (9.43) \\
pllean-cut & 170 (8.18) \\
pllean-comp7 & 159 (7.65) \\
mlleancop-nocut & 150 (7.21) \\
mlleancop-cut & 146 (7.02) \\
meson-infer & 145 (6.97) \\
metis-23 & 138 (6.64) \\
pllean-nc & 126 (6.06) \\
any & 693 (33.34) \\\bottomrule
\end{tabular}
\caption{\label{tab:newchainy} Chainy (large) MPTP2078  problems (2078 in total), 10sec)}
\end{table}

\section{Conclusion}
We have implemented an \OCaml version of the \leanCoP compact
connection prover, and the reconstruction of its proofs inside
\HOLLight. This proof-reconstruction functionality can be also used to certify in \HOLLight an arbitrary TPTP proof produced by \leanCoP, thus turning
\leanCoP into one of the few ATPs whose proofs enjoy LCF-style verification in one of the safest LCF-based systems.
The performance of the \OCaml version on the benchmarks is comparable to the
\Prolog version, while it always outperforms \Metis and \MESON,
sometimes very significantly on the relevant ITP-related benchmarks.

We provide a \HOLLight interface that is identical to the one offered
by \texttt{MESON}, namely we provide two tactics and a
rule. \texttt{LEANCOP\_TAC} and \texttt{ASM\_LEANCOP\_TAC} are given a
list of helper theorems, and then try to solve the given goal together
(or the given goal and assumptions, respectively). The \texttt{LEANCOP} rule,
given a list of helper theorems acts as a conversion, i.e., given a
term statement it tries to prove a theorem whose conclusion is
identical to that of the term. The benchmarks show that these are
likely the strongest single-step proof-reconstructing 
first-order tactics available today in any ITP system.

\acks

Supported by the Austrian Science Fund (FWF): P26201.

\bibliography{../ate11}
\bibliographystyle{abbrv}

\end{document}